\documentstyle[11pt,epsf]{article}

\newfont{\subsub}{cmr6}

\newcounter{szk}

\begin{document}

\title{\bf Quasistatically varying log-normal distribution 
in the middle scale region of Japanese land prices
}
\author{
\footnote{e-mail address: ishikawa@kanazawa-gu.ac.jp} 
Atushi Ishikawa
\\
Kanazawa Gakuin University, Kanazawa 920-1392, Japan
}
\date{}
\maketitle

\begin{abstract}
{
Employing data on the assessed value of land in 1974--2007 Japan,
we exhibit a quasistatically varying log-normal distribution in the middle scale region.
In the derivation, a Non-Gibrat's law under the detailed quasi-balance
is adopted together with two approximations. 
The resultant distribution is power-law with the varying exponent 
in the large scale region
and the quasistatic log-normal distribution 
with the varying standard deviation
in the middle scale region.
In the distribution,
not only the change of the exponent 
but also the change of the standard deviation
depends on the parameter of the detailed quasi-balance.
These results are consistently confirmed by the empirical data.
}
\end{abstract}
\begin{flushleft}
\end{flushleft}

\section{Introduction}
\label{sec-Introduction}
\indent

Log-normal distributions are frequently observed 
not only in natural phenomena but also in social ones.
For representative example, 
the probability density function $P(x)$ of personal income or
firm size $x$
is considered to obey the log-normal distribution~\cite{Gibrat}--\cite{Badger}
\begin{eqnarray}
    P_{\rm LN}(x) = \frac{1}{x \sqrt{2 \pi \sigma^2}} \exp 
        \left[-\frac{\ln^2 \left( x/\bar{x} \right)}{2 \sigma^2} \right]
    \label{LN}        
\end{eqnarray}
in the middle scale region.\footnote{
In the large scale region, the distribution of
personal income or firm size
follows power-law \cite{Pareto}. We discuss this in the next section.
In the low scale region, several distributions are proposed
(see Refs.~\cite{Yakovenko}--\cite{AIST} for instance).
We do not discuss them in this study.
}
Here $\bar{x}$ is a mean value and $\sigma^2$ is a variance.
A large number of persons or firms are included in the middle scale region.
The study of the distributions is significant.

The simplest model which describes the log-normal distribution
is the pure multiplicative stochastic process defined by
\begin{eqnarray}
    x(t+1) = R(t)~x(t)~,    
    \label{MSP}
\end{eqnarray}
where $R(t)$ is a positive random variable. 
By applying this process iteratively, we obtain
\begin{eqnarray}
    x(t) = R(t-1)~R(t-2)~\cdots~R(0)~x(0)~.  
    \label{MSP2}
\end{eqnarray}
The logarithm of this equation is
\begin{eqnarray}
    \log x(t) = \log R(t-1)+\log R(t-2)+\cdots+\log R(0)+\log x(0)~.  
    \label{MSP3}
\end{eqnarray}
If $\log x(0)$ is negligible compared to $\log x(t)$ in the limit $t \to \infty$ and
$\log R(i)$ $(i = 0, 1, \cdots, t-1)$ are independent probability variables, 
$\log x(t)$ follows the normal distribution in the limit.
As a result, this model leads the stationary log-normal distribution $P_{\rm LN} (x)$.

Eq.~(\ref{MSP}) means that the distribution of the growth rate $R(t)$
does not depend on $x(t)$.
This is known as Gibrat's law \cite{Gibrat}
that the conditional probability density function $Q(R|x_1)$ of the growth rate
is independent of the initial value $x_1$:
\begin{eqnarray}
    Q(R|x_1) = Q(R)~.
    \label{Gibrat}
\end{eqnarray}
Here $x_1$ and $x_2$ are two successive incomes, assets, sales, profits,
the number of employees and so forth.
The growth rate $R$ is defined as $R=x_2/x_1$ and
$Q(R|x_1)$ as 
\begin{eqnarray}
    Q(R|x_1) = \frac{P_{1 R} (x_1, R) }{ P(x_1)}
    \label{define}
\end{eqnarray}
by using the probability density function $P(x_1)$ and the joint probability
density function $P_{1 R}(x_1, R)$.

As far as firm sizes in the middle scale region, however, it is reported that
the growth rate distributions do not follow the Gibrat's law (\ref{Gibrat}) 
(see Refs.~\cite{Stanley1}--\cite{Aoyama} for instance).
The log-normal distribution in the middle scale region
cannot be explained by the pure multiplicative stochastic process model (\ref{MSP}).
Instead,
we have shown that the log-normal distribution can be derived~\cite{Ishikawa3}
by using no model such as the pure multiplicative stochastic 
process.
In the derivation, two laws are employed which are observed 
in profits data of Japanese firms.
One is the law of detailed balance
which represents symmetry in a stable economy \cite{FSAKA}.
The other is a Non-Gibrat's law 
which describes a statistical dependence 
in the growth rate of the past value~\cite{Ishikawa3}.

In Ref.~\cite{Ishikawa2007},
the Non-Gibrat's law and the static log-normal distribution
in the middle scale region
are uniquely derived from the detailed balance.
The derivation has been confirmed by the empirical data analysis.
In this study, 
we extend the derivation
by replacing the detailed balance with the detailed quasi-balance proposed 
in Ref.~\cite{Ishikawa1} to 
derive a log-normal distribution
in the quasistatic system. 
By this procedure, the log-normal distribution is described as quasistatic.
The derivation is consistently confirmed employing data on the assessed value of land in 1974--2007 Japan.

\section{Static log-normal distribution under the detailed balance}
\label{Static log-normal distribution under the detailed balance}
\indent

In this section, we briefly review the study in Ref.~\cite{Ishikawa2007}.
As an equilibrium system,
we investigate profits data of Japanese firms in 2003, 2004 and 2005
which are available on the database
``CD Eyes 50" published in 2005 and 2006 by TOKYO SHOKO RESEARCH, LTD.~\cite{TSR}.

\begin{figure}[htb]
 \centerline{\epsfxsize=0.75\textwidth\epsfbox{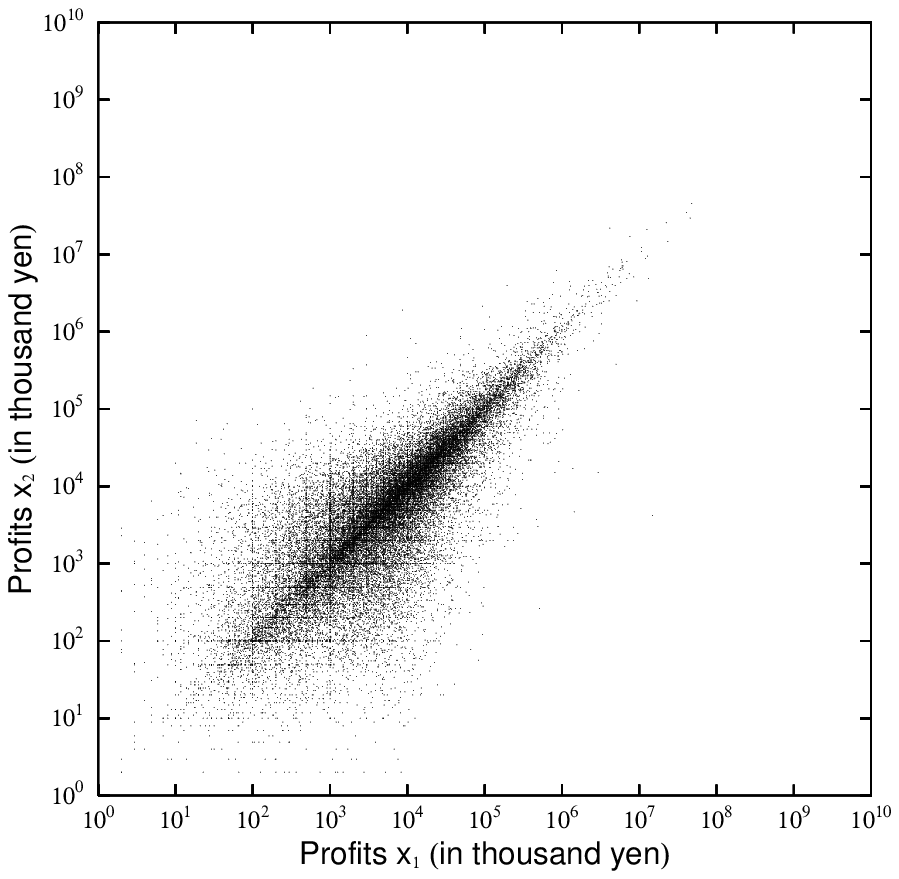}}
 \caption{The scatter plot of all firms in the database,
 the profits of which in 2004 ($x_1$) and 2005 ($x_2$) exceeded $0$,
 $x_1 >0$ and $x_2 > 0$.
 The number of firms is ``232,497 ".}
 \label{logProfit0405DB}
\vspace{1.5cm}
 \centerline{\epsfxsize=0.75\textwidth\epsfbox{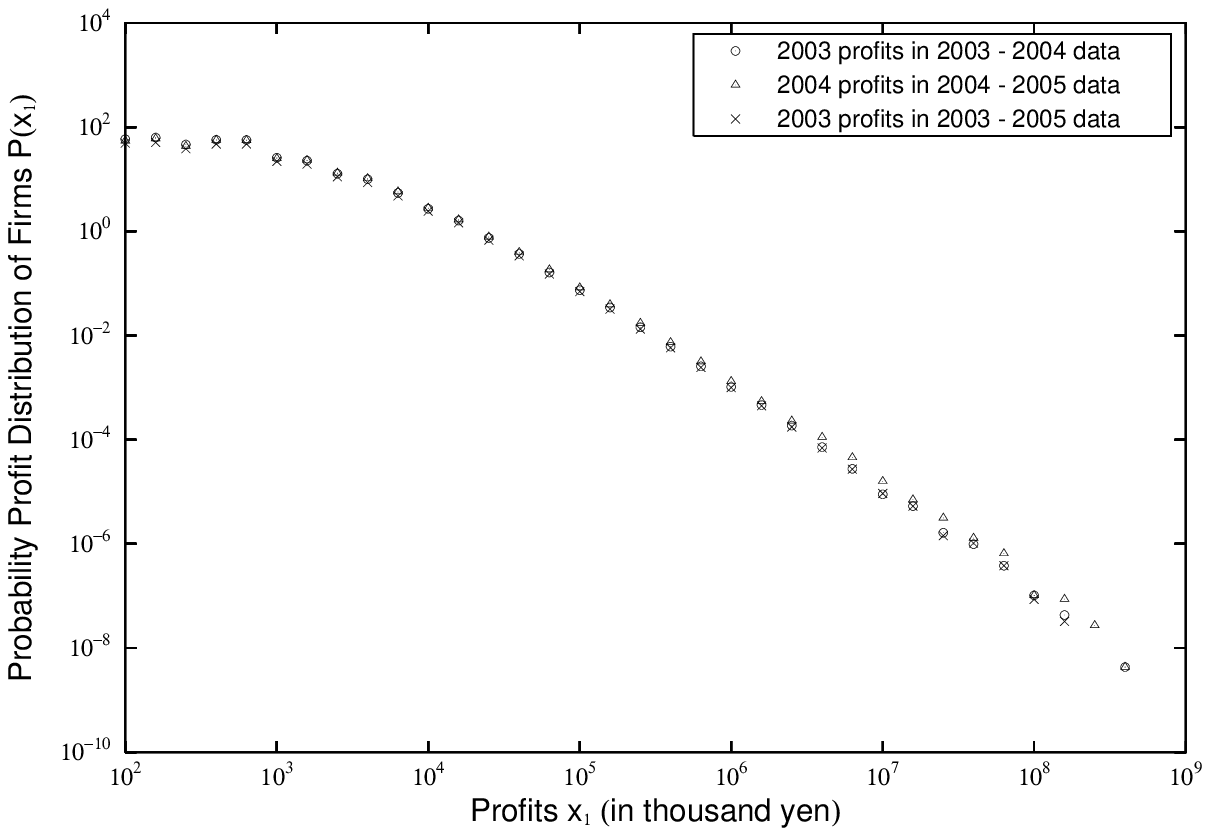}}
 \caption{Probability density functions of profits 
in the database.
In the large scale region, Pareto's law is observed. Each
Pareto index is estimated to be nearly $1$.
}
 \label{DistributionData}
\end{figure}

Figure \ref{logProfit0405DB} shows 
the joint probability density function $P_{1 2} (x_1, x_2)$
of all firms in the database,
the profits of which in 2004 ($x_1$) and 2005 ($x_2$) exceeded $0$,
$x_1 >0$ and $x_2 > 0$.
The number of firms is ``232,497". From Fig.~\ref{logProfit0405DB},
we approximately confirm the detailed balance
which is time-reversal symmetry ($x_1 \leftrightarrow x_2$) of 
$P_{1 2} (x_1, x_2)$ \cite{FSAKA}:\footnote{
Similarly, we approximately confirm the detailed balance in the joint probability density
functions of the profits in 2003--2004 and 2003--2005.}
\begin{eqnarray}
    P_{1 2}(x_1, x_2) = P_{1 2}(x_2, x_1)~.
    \label{Detailed balance}
\end{eqnarray}

Figure \ref{DistributionData} shows probability density functions of profits 
in the database.
The distributions are almost stable and the following power-law is observed
in the large scale region
\begin{eqnarray}
    P(x) = C x^{-\mu-1}~~~~{\rm for }~~~~x > x_0~,
    \label{Pareto}
\end{eqnarray}
where $x_0$ is a certain threshold.
This power-law is called Pareto's law \cite{Pareto}
and the exponent $\mu$ is named Pareto index.
Notice that the Pareto's law does not hold below the threshold $x_0$.
The purpose of this section is to exhibit the distribution in the middle scale region
under the detailed balance (\ref{Detailed balance}).

\begin{figure}[htb]
 \begin{minipage}[htb]{0.49\textwidth}
  \epsfxsize = 1.0\textwidth
  \epsfbox{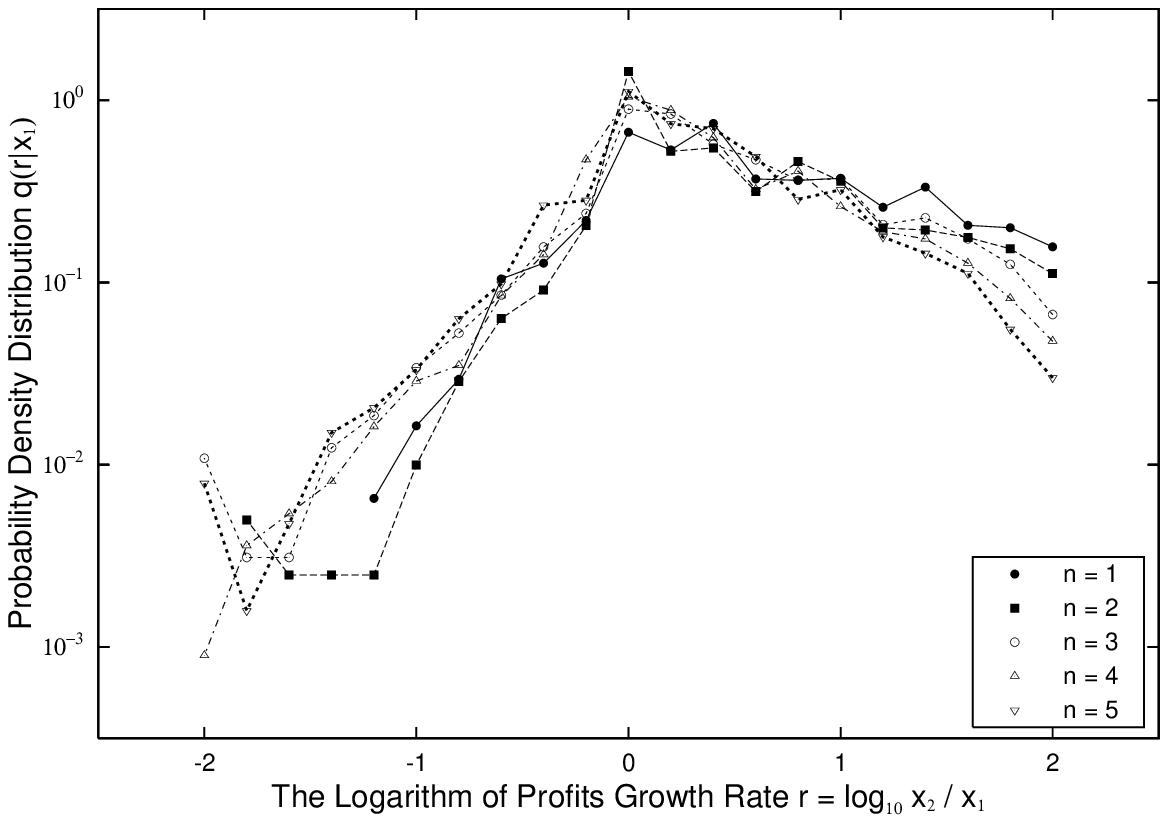}
 \end{minipage}
 \hfill
 \begin{minipage}[htb]{0.49\textwidth}
  \epsfxsize = 1.0\textwidth
  \epsfbox{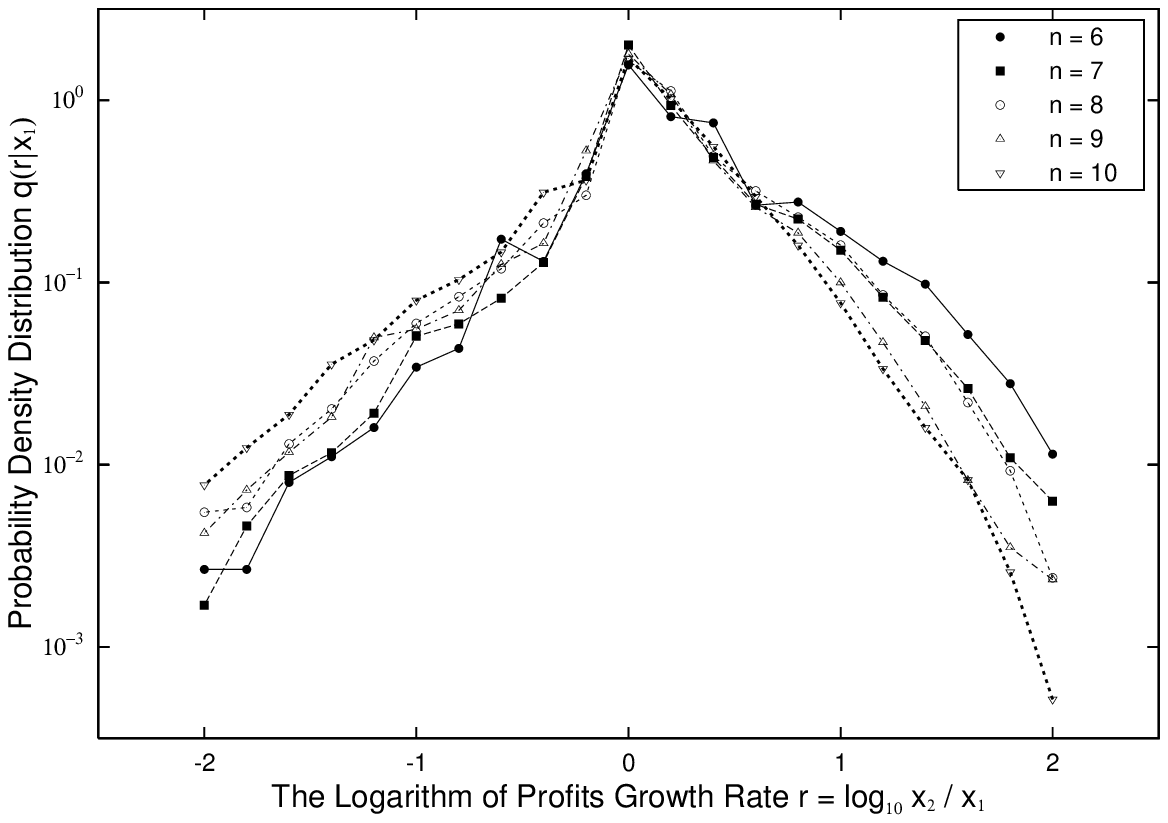}
 \end{minipage}
 \begin{minipage}[htb]{0.49\textwidth}
  \epsfxsize = 1.0\textwidth
  \epsfbox{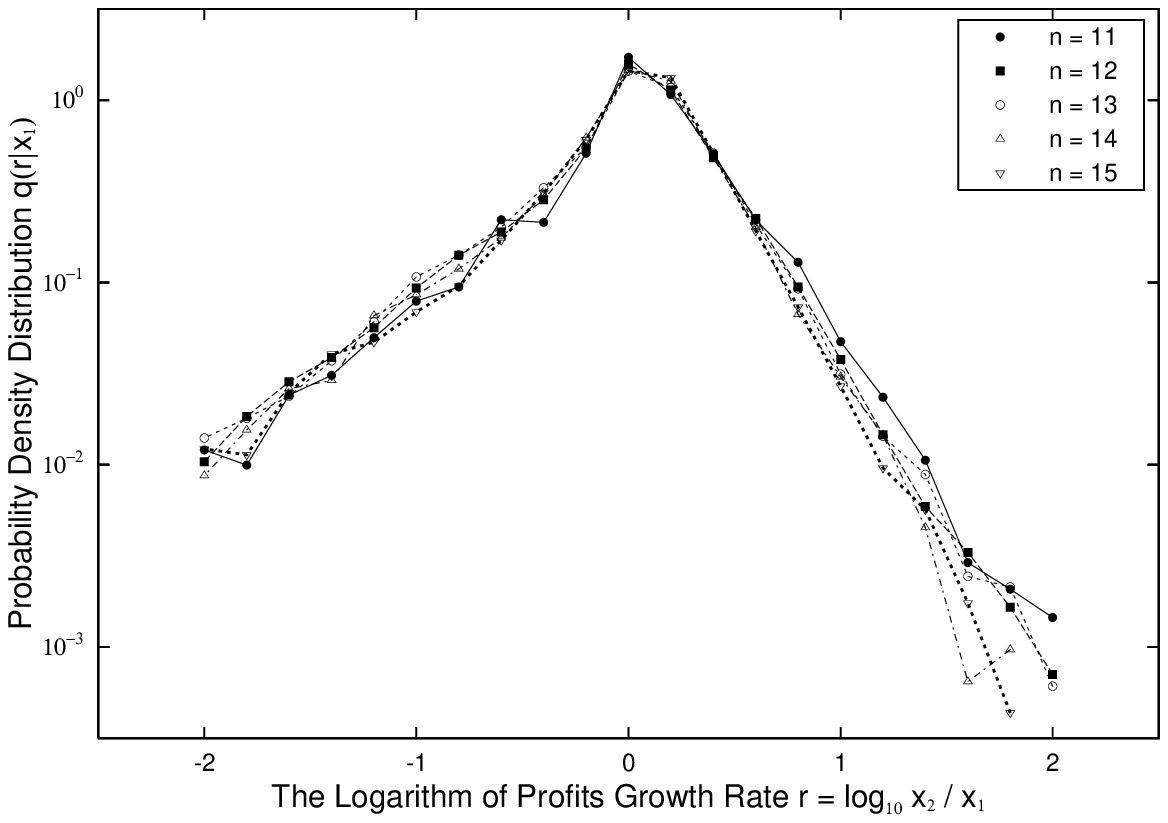}
 \end{minipage}
 \hfill
 \begin{minipage}[htb]{0.49\textwidth}
  \epsfxsize = 1.0\textwidth
  \epsfbox{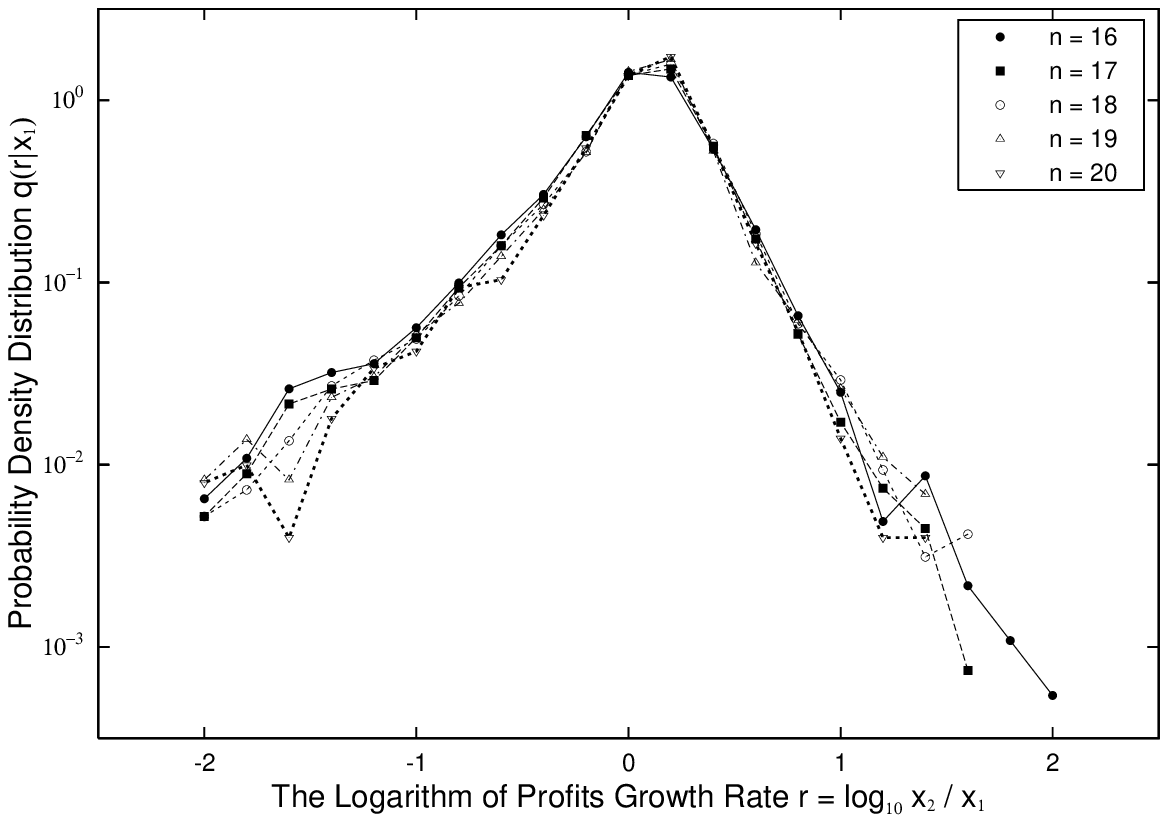}
 \end{minipage}
 \caption{Conditional probability density distributions $q(r|x_1)$ of the log profits growth rate 
 $r = \log_{10} x_2/x_1$ 
 from $2004$ to $2005$ for $n=1, 2, \cdots, 20$.}
 \label{ProfitGrowthRate0405}
\end{figure}

In order to identify a statistical dependence 
in the growth rate of the past value,
we examine the probability density function of the profits growth rate in 2004--2005.
In the database, we divide the range of $x_1$ into logarithmically equal bins
as $x_1 \in 4 \times [10^{1+0.2(n-1)},10^{1+0.2n}]$ thousand yen with $n=1, 2, \cdots, 20$.
Figure \ref{ProfitGrowthRate0405} shows the conditional probability density functions for $r = \log_{10} R$
in the case of $n=1, \cdots, 5$, $n=6, \cdots, 10$, $n=11, \cdots, 15$
and $n=16, \cdots, 20$.
The number of firms in Fig.~\ref{ProfitGrowthRate0405}
is ``$20,669$", ``$89,064$", ``$88,651$" and ``$26,887$", respectively.

From Fig.~\ref{ProfitGrowthRate0405},
we approximate $\log_{10} q(r|x_1)$ by linear functions of $r$:
\begin{eqnarray}
    \log_{10}q(r|x_1)&=&c(x_1)-t_{+}(x_1)~r~~~~~{\rm for}~~r > 0~,
    \label{approximation1}\\
    \log_{10}q(r|x_1)&=&c(x_1)+t_{-}(x_1)~r~~~~~{\rm for}~~r < 0~.
    \label{approximation2}
\end{eqnarray}
Here $q(r|x_1)$ is the conditional probability density function for $r$,
which is related to that for $R$ by $\log_{10} Q(R|x_1) = \log_{10}q(r|x_1) - r - \log_{10}(\ln 10)$.
These approximations (\ref{approximation1})--(\ref{approximation2})
are expressed as tent-shaped exponential forms as follows:
\begin{eqnarray}
    Q(R|x_1)&=&d(x_1)~R^{-t_{+}(x_1)-1}~~~~~{\rm for}~~R > 1~,
    \label{tent-shaped1}\\
    Q(R|x_1)&=&d(x_1)~R^{+t_{-}(x_1)-1}~~~~~{\rm for}~~R < 1~,
    \label{tent-shaped2}
\end{eqnarray}
where $d(x_1)=10^{c(x_1)}/{\ln 10}$.
Furthermore, Fig.~\ref{ProfitGrowthRate0405} shows that
the dependence of $c(x_1)$ on $x_1$ is negligible for $n = 9, \cdots, 20$.
We assess the validity of these approximations against the results.

\begin{figure}[htb]
 \begin{minipage}[htb]{0.49\textwidth}
  \epsfxsize = 1.0\textwidth
  \epsfbox{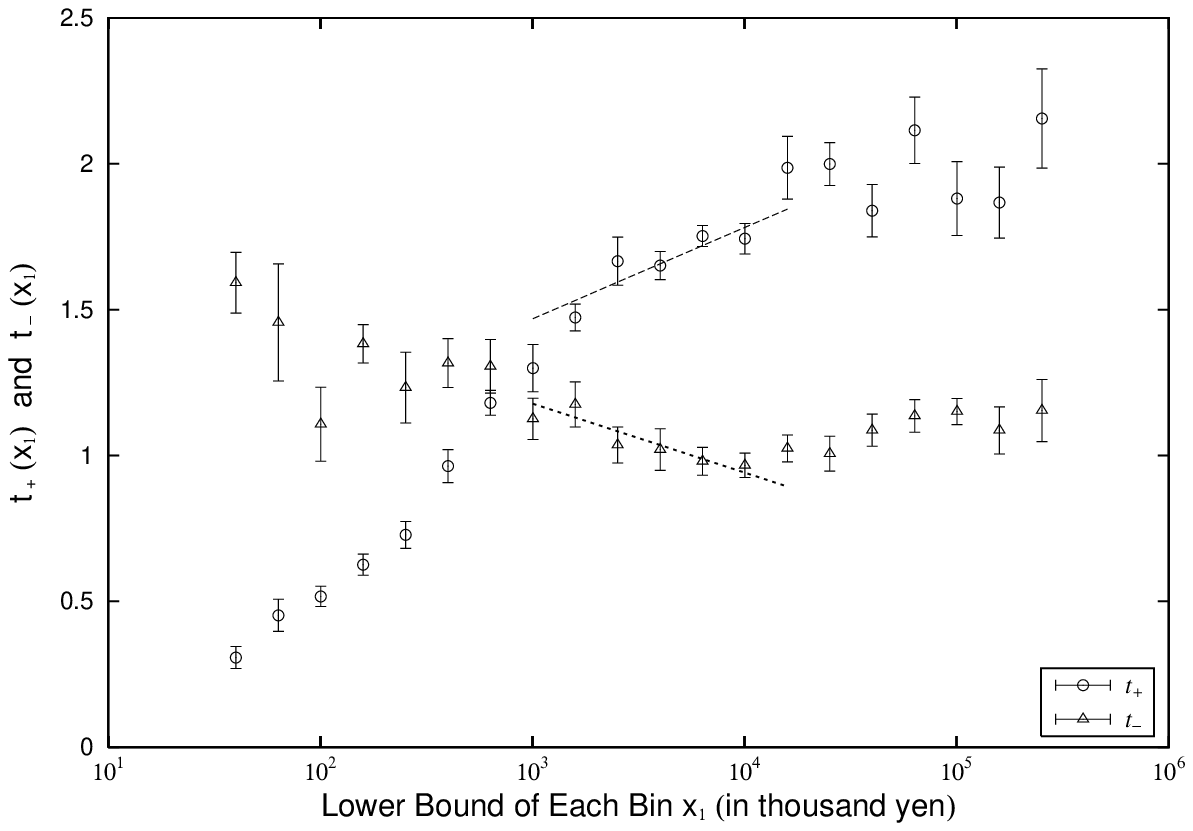}
 \caption{The relation between the lower bound of each bin $x_1$ and $t_{\pm}(x_1)$
 with respect to the profits growth rate from $2004$ to $2005$.
 From the left, each data point represents $n=1, 2, \cdots, 20$.}
 \label{eGibrat0405}
 \end{minipage}
 \hfill
 \begin{minipage}[htb]{0.49\textwidth}
  \epsfxsize = 1.0\textwidth
  \epsfbox{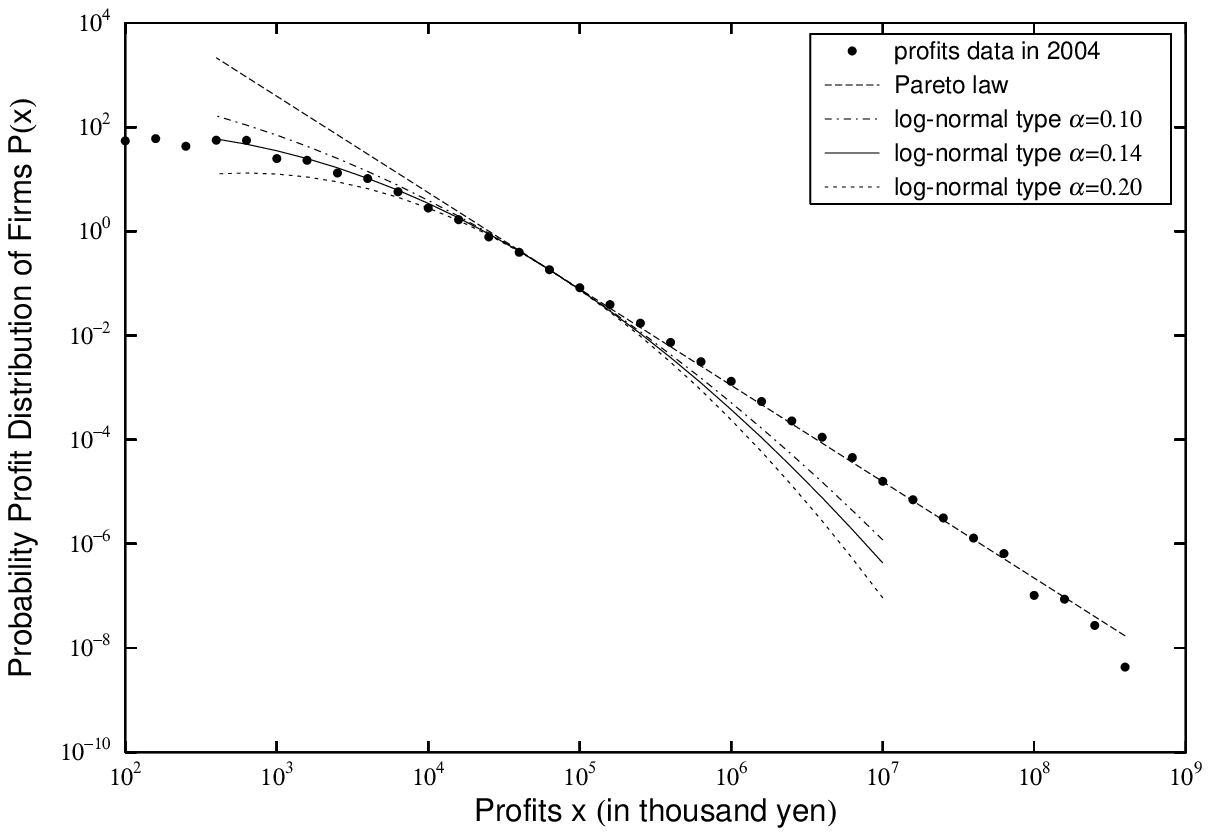}
 \caption{The probability distribution function of profits $P(x_1)$ for firms,
 the profits of which in 2004 ($x_1$) and 2005 ($x_2$) exceeded $0$,
 $x_1 >0$ and $x_2 > 0$.
 }
 \vspace{6.5mm}
 \label{Distribution0405-04}
 \end{minipage}
\end{figure}

Figure~\ref{eGibrat0405} represents
the dependence of $t_{\pm}(x_1)$ on the lower bound of each bin
$x_1 = 4 \times 10^{1+0.2(n-1)}$.
For $n=17, \cdots, 20$, $t_{\pm}(x_1)$ hardly responds to $x_1$.
This means that the Gibrat's law (\ref{Gibrat}) holds only 
in the large scale region of profits ($x > x_0$).\footnote{
Fujiwara et al.~\cite{FSAKA} prove that the Pareto's law (\ref{Pareto}) is derived
from the Gibrat's law (\ref{Gibrat}) valid only in the large scale region
and the detailed balance (\ref{Detailed balance}).
In the derivation, linear approximations (\ref{approximation1})--(\ref{approximation2})
need not be assumed.
}
In contrast, $t_{+}(x_1)$ linearly increases and $t_{-}(x_1)$ linearly decreases 
symmetrically with $\log_{10} x_1$ for $n = 9, \cdots, 13$. From
Fig.~\ref{eGibrat0405}, the slops are described as~\cite{Ishikawa3} 
\begin{eqnarray}
    t_{\pm}(x_1)=t_{\pm}(x_0) \pm \alpha~\ln \frac{x_1}{x_0}~.
    \label{t0}
\end{eqnarray}
The parameters are estimated as follows: 
$\alpha \sim 0$ for $x_1 > x_0$, 
$\alpha \sim 0.14$ for $x_{\rm min} < x_1 < x_0$,\footnote{
Here, a constant parameter $\alpha$ takes different values in two regions.
This is not an exact procedure.
However, for firms which are in the large scale region in both years ($x_1 > x_0$ and $x_2 > x_0$) 
or in the middle scale one ($x_{\rm min} < x_1 < x_0$ and $x_{\rm min} < x_2 < x_0$),
this procedure is exact. In the database,
most firms stay in the same region.
This parameterization is, therefore, approximately valid for
describing the probability density function $P(x_1)$.
This is confirmed in Fig.~\ref{Distribution0405-04}.}
$x_0 = 4 \times 10^{1+0.2(17-1)} \sim 63,000$ thousand yen and
$x_{\rm min} = 4 \times 10^{1+0.2(9-1)} \sim 1,600$ thousand yen.
Notice that 
approximations (\ref{approximation1})--(\ref{approximation2})
uniquely fix
the expression of $t_{\pm}(x_1)$
under the detailed balance~\cite{Ishikawa2007}.
This derivation is included in the proof in the next section.
We call Eqs.~(\ref{tent-shaped1})--(\ref{t0}) a Non-Gibrat's law.

For $n = 9, \cdots, 20$, the dependence of $d(x_1)$ on $x_1$ is negligible.
In this case, the Non-Gibrat's law determines the probability density function of profits
as follows:
\begin{eqnarray}
    P(x_1) = C {x_1}^{-\left(\mu+1\right)}~e^{-\alpha \ln^2 \frac{x_1}{x_0}}
    ~~~~~{\rm for}~~x_1 > x_{\rm min}~,
    \label{HandM}
\end{eqnarray}
where $t_{+}(x_0) - t_{-}(x_0) \sim \mu$~\cite{Ishikawa0}.
This is the power-law in the large scale region ($x_1 > x_0$)
and the log-normal distribution in the middle scale one 
($x_{\rm min} < x_1 < x_0$).
The relations between parameters $\sigma^2$, $\bar{x}$ in Eq.~(\ref{LN})
and $\alpha$, $\mu$, $x_0$ are given by
    $\alpha = \frac{1}{2 \sigma^2}$, $\mu = \frac{1}{\sigma^2} \ln \frac{x_0}{\bar{x}}$.
Figure~\ref{Distribution0405-04} shows that
the distribution (\ref{HandM}) fits with the empirical data consistently.
Notice that the distribution cannot fit with the empirical data,
if $\alpha$ is different from the value estimated in Fig.~\ref{eGibrat0405}
($\alpha = 0.10$ or $\alpha = 0.20$ for instance).\footnote{
These empirical data analyses are not restricted in the single term 2004--2005.
Similar data analyses are checked in 2003--2004 or 2003--2005.
}

\section{Quasistatic log-normal distribution under the detailed quasi-balance}
\label{Quasi-static log-normal distribution under the detailed quasi-balance}
\indent

In the previous section, 
the log-normal distribution in the middle scale region
is exhibited by a Non-Gibrat's law under the detailed balance.
The resultant profits distribution is empirically confirmed 
in data analyses of Japanese firms in 2003--2004, 2004--2005 and 2003--2005.
The profits distribution (\ref{HandM}) is static, 
because the derivation is based on the detailed balance (\ref{Detailed balance})
which is static time-reversal symmetry observed in the system
(Fig.~\ref{logProfit0405DB}).

In the case that an economy is not stable,
the detailed balance should be extended to describe the state.
In Ref.~\cite{Ishikawa1},
we have derived Pareto's law with annually varying Pareto index
under the detailed quasi-balance:
\begin{eqnarray}
    P_{1 2}(x_1, x_2) 
    = P_{1 2}( \left( \frac{x_2}{a} \right)^{1/{\theta}}, a~{x_1}^{\theta})~.
    \label{Detailed quasi-balance}
\end{eqnarray}
It is assumed that, in an ideal quasistatic system,
the joint probability density function
has ``$a~{x_1}^{\theta} \leftrightarrow x_2$" symmetry
where $\theta$ is a slope of a regression line: 
\begin{eqnarray}
    \log_{10} x_2 = \theta~\log_{10} x_1 + \log_{10} a~.
    \label{Line}
\end{eqnarray}
The detailed balance (\ref{Detailed balance}) 
has the special symmetry $\theta = a = 1$.
Because the detailed quasi-balance (\ref{Detailed quasi-balance})
is imposed on the system,
$\theta$ is related to $a$ as follows:
\begin{eqnarray}
    \theta = 1 - \frac{2}{\Gamma} \log_{10} a~.
    \label{Gamma0}  
\end{eqnarray}
Here $10^{\Gamma}$ is a sufficient large value compared to the upper bound 
in which $\theta$ and $a$ are estimated.

In Ref.~\cite{Ishikawa1},
these results have been empirically confirmed 
by employing data on the assessed value of land~\cite{Kaizoji}--\cite{Web} 
in 1983--2005.
In the derivation,
under the detailed quasi-balance,
we have used the Gibrat's law (\ref{Gibrat}) valid only in the large scale region
without linear approximations (\ref{approximation1})--(\ref{approximation2}).
The purpose of this section is to show that the 
approximations uniquely fix
a Non-Gibrat's law 
even under the detailed quasi-balance (\ref{Detailed quasi-balance}).
After that, we identify the quasistatic distribution not only in the large scale region
but also in the middle scale one.

By using the relation $P_{1 2}(x_1, x_2)dx_1 dx_2 = P_{1 R}(x_1, R)dx_1 dR$
under the change of variables $(x_1, x_2)$ $\leftrightarrow  (x_1, R)$,
these two joint probability density functions are related to each other
\begin{eqnarray}
    P_{1 R}(x_1, R) = a~{x_1}^{\theta}~P_{1 2}(x_1, x_2)~,
\end{eqnarray}
where we use a modified growth rate $R= x_2/(a~{x_1}^{\theta})$~. From this relation,
the detailed quasi-balance (\ref{Detailed quasi-balance}) is rewritten as 
\begin{eqnarray}
    P_{1 R}(x_1, R) = R^{-1} P_{1 R}(\left(\frac{x_2}{a}\right)^{1/\theta}, R^{-1})~.
\end{eqnarray}
Substituting $P_{1 R}(x_1, R)$ for $Q(R|x_1)$ defined in Eq.~(\ref{define}),
the detailed quasi-balance is reduced to be
\begin{eqnarray}
    \frac{P(x_1)}{P(\left(x_2/a \right)^{1/\theta})} 
    = \frac{1}{R} \frac{Q(R^{-1}|\left(x_2/a \right)^{1/\theta})}{Q(R|x_1)}~.
    \label{}
\end{eqnarray}

In the quasistatic system, we also assume 
that $Q(R|x_1)$ follows the tent-shaped exponential forms 
(\ref{tent-shaped1})--(\ref{tent-shaped2}).\footnote{
For instance, the tent-shaped exponential forms are observed 
in data on the assessed value of land
in Japan~\cite{Kaizoji}.}
Under the approximations, the detailed quasi-balance is expressed as
\begin{eqnarray}
    \frac{\tilde{P}(x_1)}{\tilde{P}(\left(x_2/a \right)^{1/\theta})} 
    = R^{~t_{+}(x_1) - t_{-}(\left(x_2/a \right)^{1/\theta}) + 1}~
    \label{Detailed quasi-balance 2}
\end{eqnarray}
for $R>1$. Here we use the notation $\tilde{P}(x)=P(x) d(x)$.
By setting $R=1$ after
differentiating Eq.~(\ref{Detailed quasi-balance 2}) with respect to $R$, 
the following differential equation is obtained:
\begin{eqnarray}
    \theta~\Bigl[t_{+}(x) - t_{-}(x) + 1 \Bigr] \tilde{P}(x) + x \tilde{P}~{'}(x) = 0~,
    \label{DE}
\end{eqnarray}
where $x$ denotes $x_1$.
The same equation is obtained for $R<1$.

Similarly, from the second and third derivatives of Eq.~(\ref{Detailed quasi-balance 2}),
the following differential equations are obtained:
\begin{eqnarray}
    {t_{+}}^{'}(x)+{t_{-}}^{'}(x)=0~,~~~
    {t_{+}}^{'}(x)+x~{t_{+}}^{''}(x)=0~.   
\end{eqnarray}
The solutions are uniquely fixed as
\begin{eqnarray}
    t_{\pm}(x)=t_{\pm}(x_0) \pm \alpha~\ln \frac{x}{x_0}~.
    \label{t}
\end{eqnarray}
This is the same expression under the detailed balance.

By using the Non-Gibrat's law (\ref{tent-shaped1})--(\ref{tent-shaped2}), (\ref{t})
and the differential equation (\ref{DE}),
probability density functions $P_1(x_1)$, $P_2(x_2)$ are also uniquely reduced to
\begin{eqnarray}
    P_1(x_1) &=& C_1 ~{x_1}^{-\mu_1-1}~
        \exp\left[-\theta~\alpha \ln^2 \frac{x_1}{x_0} \right],~~~
    \label{HandM1}\\
    P_2(x_2) &=& C_2 ~{x_2}^{-\mu_2-1}~
        \exp\left[-\theta~\alpha \ln^2 \frac{\left(x_2/a \right)^{1/\theta}}{x_0}\right]~
    \label{HandM2}
\end{eqnarray}
with
\begin{eqnarray}      
    \frac{\mu_1+1}{\mu_2+1} &=& \theta~
    \label{ratio1}  
\end{eqnarray}
in the large or middle scale
region where the dependence of $d(x_1)$ on $x_1$ is negligible.

Here we consider two log-normal distributions
in the middle scale region:
\begin{eqnarray}
    P_{\rm LN_1}(x_1) = \frac{1}{x_1 \sqrt{2 \pi {\sigma_1}^2}} \exp 
        \left[-\frac{\ln^2 \left( x_1/\bar{x}_1 \right)}{2 {\sigma_1}^2} \right]~,
    \label{LN1}\\
    P_{\rm LN_2}(x_2) = \frac{1}{x_2 \sqrt{2 \pi {\sigma_2}^2}} \exp 
        \left[-\frac{\ln^2 \left( x_2/\bar{x}_2 \right)}{2 {\sigma_2}^2} \right]~.
    \label{LN2}        
\end{eqnarray}
By comparing Eqs.~(\ref{HandM1})--(\ref{HandM2}) to (\ref{LN1})--(\ref{LN2}), we identify
    $\theta \alpha = \frac{1}{2 {\sigma_1}^2}$,
    $\mu_1 = \frac{1}{{\sigma_1}^2} \ln \frac{x_0}{\bar{x}_1}$,
    $\frac{\alpha}{\theta} = \frac{1}{2 {\sigma_2}^2}$ and
    $\mu_2 = \frac{1}{{\sigma_2}^2} \ln \frac{a {x_0}^{\theta}}{\bar{x}_1}$.
Consequently, the relation between $\sigma_1$, $\sigma_2$ and $\theta$ is expressed as
\begin{eqnarray}
    \frac{\sigma_2}{\sigma_1} = \theta~.
    \label{ratio2} 
    \end{eqnarray}
This is the equation which quasistatic log-normal distributions satisfy.

\section{Data Analysis}
\label{Data Analysis}
\indent

In this section, we confirm the results in the previous section 
employing data on the assessed value of land in 1974--2007 Japan.
In Japan, land is a very important asset which is distinguished from buildings.
The assessed value of land indicates the standard land prices evaluated by 
Ministry of Land, Infrastructure and Transport Japan.
The investigation is undertaken on each piece of land assessed once a year
according to the posted land price system from 1970.\footnote{
We exclude data in 1970--1973, the number of which is insufficient.}

\begin{figure}[htb]
 \begin{minipage}[htb]{0.49\textwidth}
  \epsfxsize = 1.0\textwidth
  \epsfbox{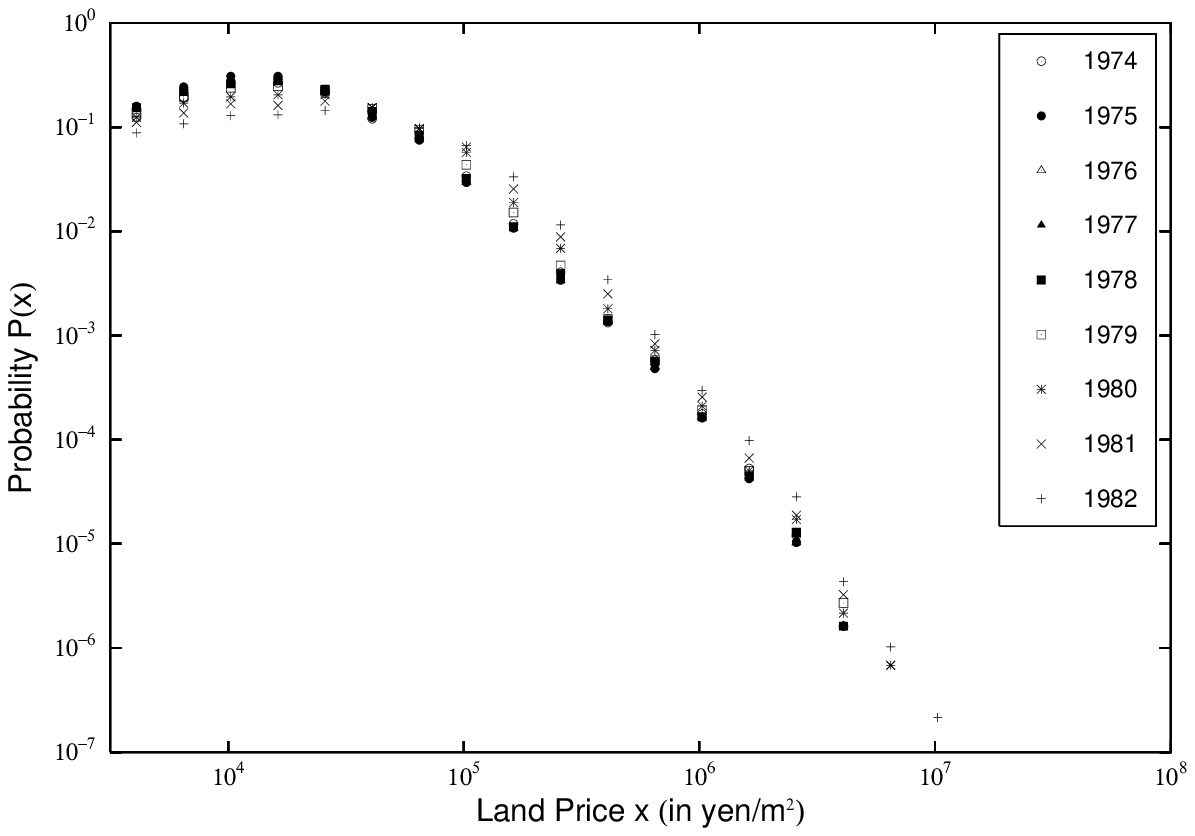}
 \end{minipage}
 \hfill
 \begin{minipage}[htb]{0.49\textwidth}
  \epsfxsize = 1.0\textwidth
  \epsfbox{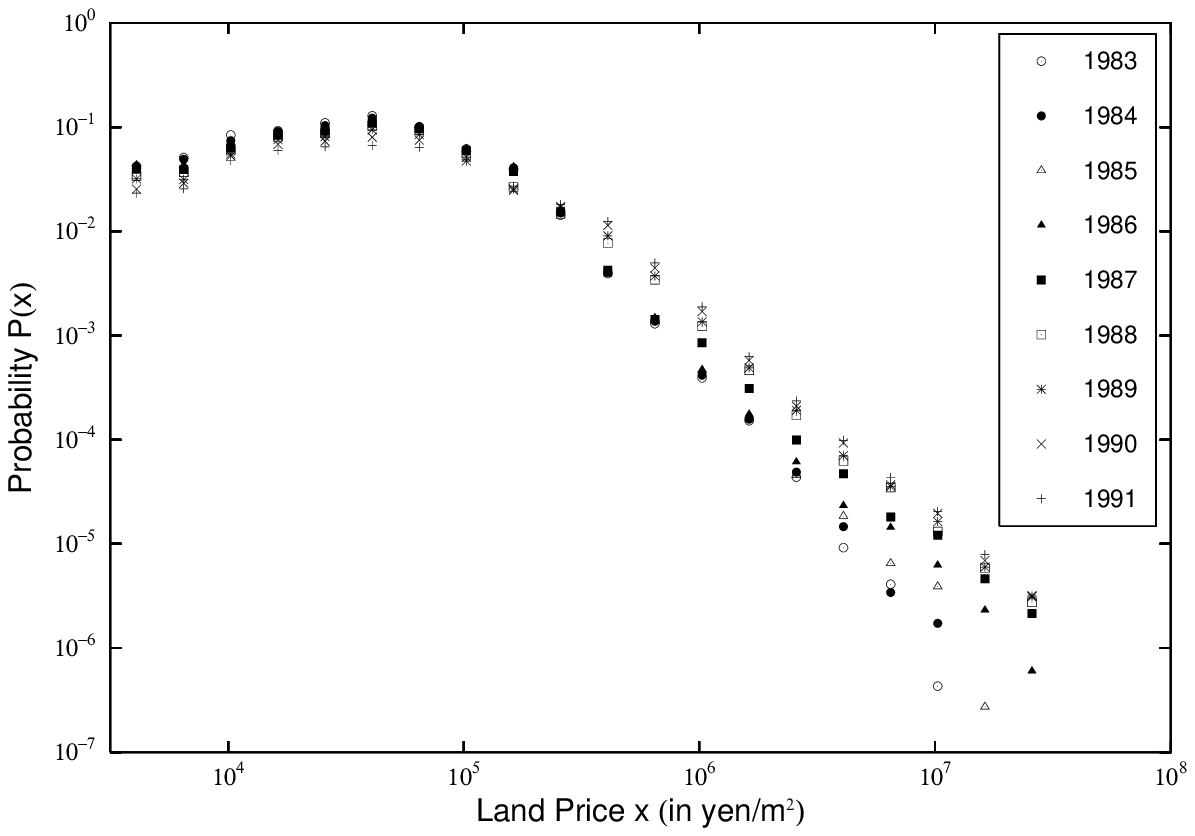}
 \end{minipage}
 \begin{minipage}[htb]{0.49\textwidth}
  \epsfxsize = 1.0\textwidth
  \epsfbox{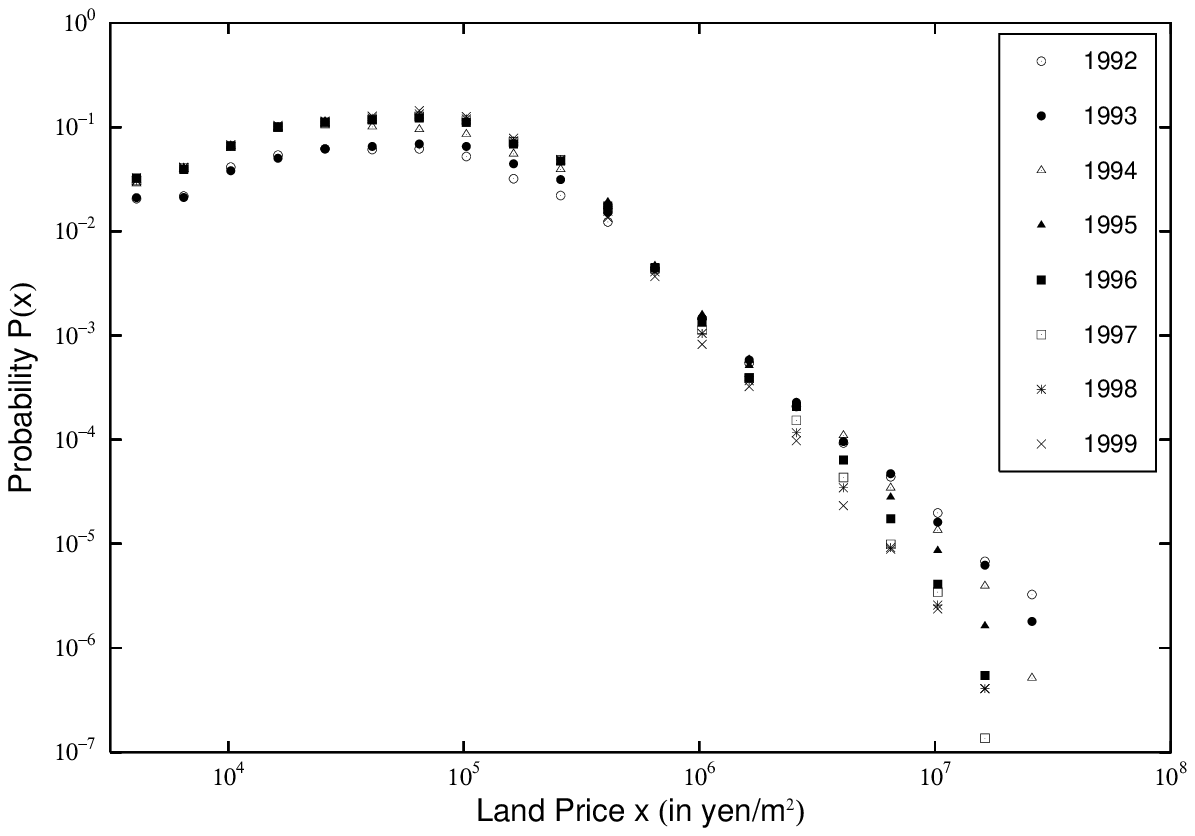}
 \end{minipage}
 \hfill
 \begin{minipage}[htb]{0.49\textwidth}
  \epsfxsize = 1.0\textwidth
  \epsfbox{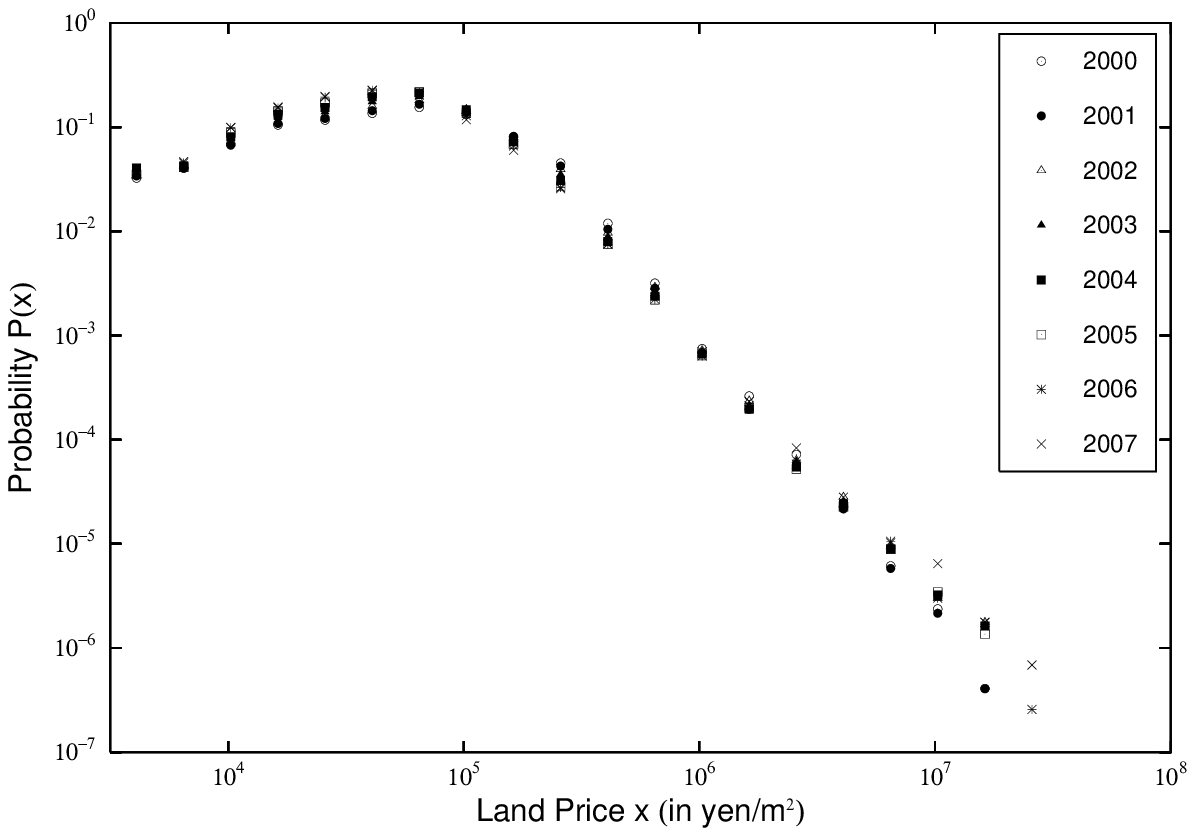}
 \end{minipage}
 \caption{Probability distributions of the assessed value of land in 1974--2007.
 In each figure, data points are equally spaced in logarithm of land price.
 }
 \label{DistPart}
\end{figure}

The probability distribution functions of land prices are shown 
in Fig.~\ref{DistPart}. 
The number of data points is ``14,570", ``15,010", ``15,010", ``15,010",
``15,580", ``16,480", ``17,030", ``17,380" and ``17,600" in 1974--1982, respectively.
In 1983--1991,
``16,975", ``16,975", ``16,975", ``16,635",
``16,635", ``16,820", ``16,840", ``16,865" and ``16,892", respectively.
In 1992--1999,
``17,115", ``20,555", ``26,000", ``30,000",
``30,000", ``30,300", ``30,600" and ``30,800", respectively.
In 2000--2007,
``31,000", ``31,000", ``31,520", ``31,866",
``31,866", ``31,230", ``31,230" and ``30,000", respectively.

\begin{figure}[hbp]
 \centerline{\epsfxsize=0.75\textwidth\epsfbox{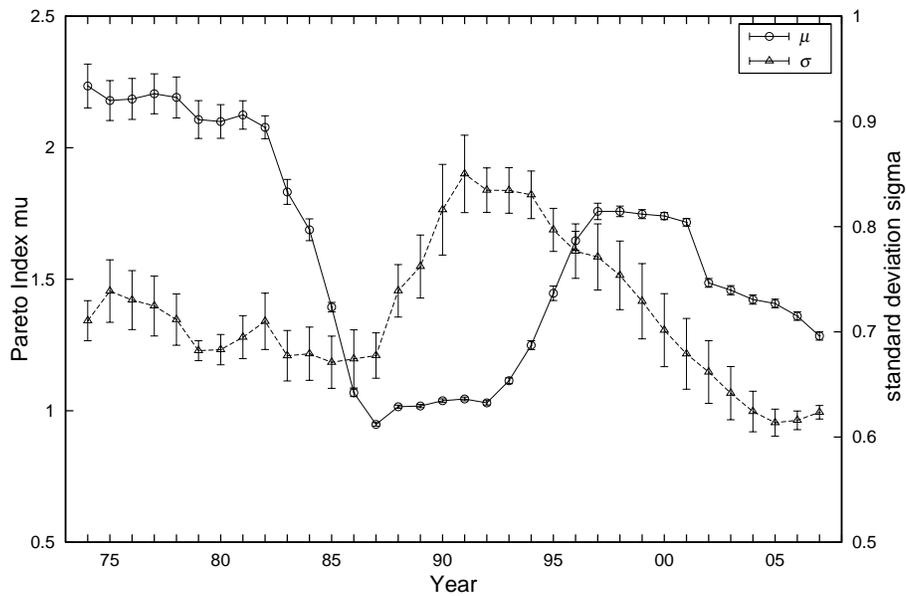}}
 \caption{Annual change of Pareto index $\mu$ and the standard deviation $\sigma$ 
 of the log-normal distribution
 from 1974 to 2007.
 }
 \label{VaryingParetoIndex2}
\end{figure}

In each figure, the power-law is observed in the large scale region.
In addition, Pareto index $\mu$ varies annually 
and changes significantly before and after bubble years (1986--1991).
This is represented in Fig.~\ref{VaryingParetoIndex2}
where each Pareto index $\mu$ is estimated in the range of land prices
from $2 \times 10^5$ to $10^7$ ${\rm yen}/{\rm m}^2$.
In Fig.~\ref{DistPart},
the log-normal distribution is also observed in each middle scale region.
The standard deviation $\sigma$ varies annually
and changes significantly in bubble years.
This is also represented in Fig.~\ref{VaryingParetoIndex2}
where each standard deviation $\sigma$ is estimated in the range 
from $5 \times 10^3$ to $3.17 \times 10^5$ ${\rm yen}/{\rm m}^2$.

\begin{figure}[htb]
 \centerline{\epsfxsize=0.75\textwidth\epsfbox{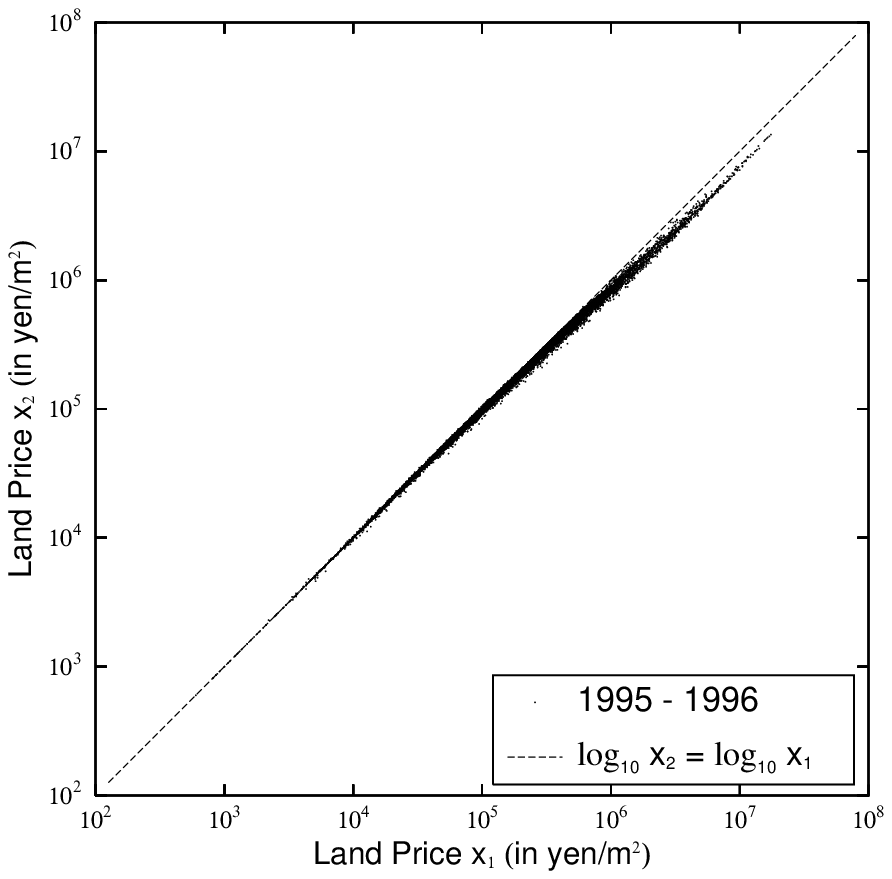}}
 \caption{The scatter plot of all pieces of land assessed in the database,
 the values of which in 1995 ($x_1$) and 1996 ($x_2$) exceeded $10^2 ~{\rm yen}/{\rm m}^2$.
 The number of data points is ``29,590''.}
 \label{95vs96}
\end{figure}

\begin{figure}[htb]
 \centerline{\epsfxsize=0.75\textwidth\epsfbox{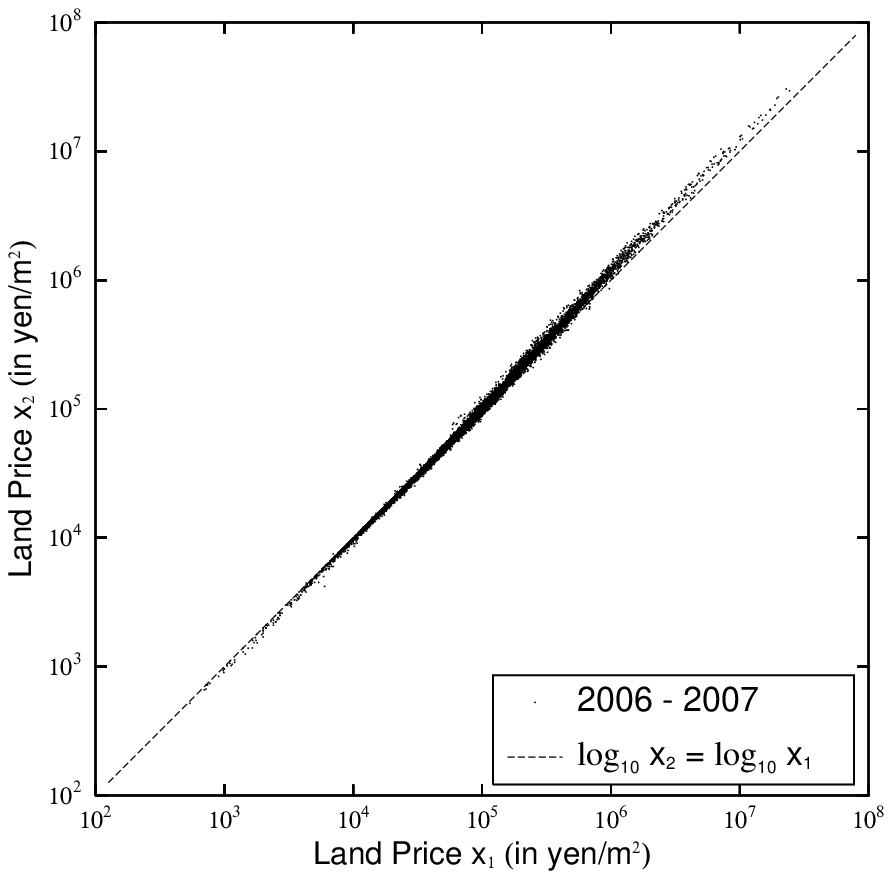}}
 \caption{The scatter plot of all pieces of land assessed in the database,
 the values of which in 1995 ($x_1$) and 1996 ($x_2$) exceeded $10^2 ~{\rm yen}/{\rm m}^2$.
 The number of data points is ``29,692''.}
 \label{06vs07}
\end{figure}

\begin{figure}[htb]
 \centerline{\epsfxsize=0.75\textwidth\epsfbox{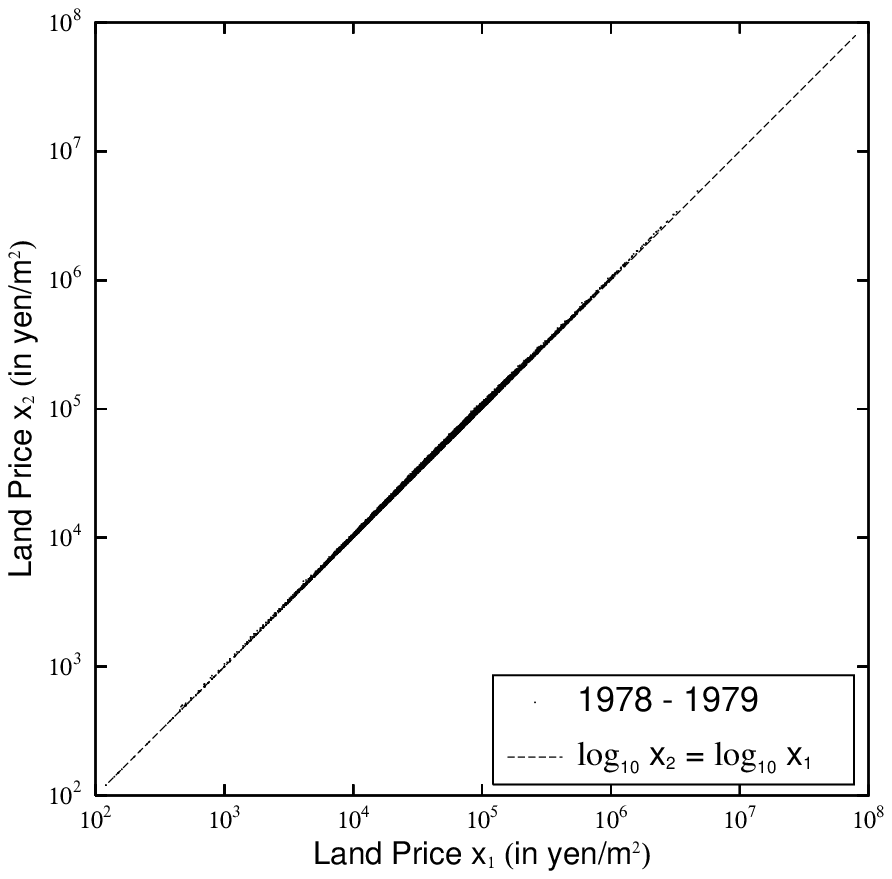}}
 \caption{The scatter plot of all pieces of land assessed in the database,
 the values of which in 1978 ($x_1$) and 1979 ($x_2$) exceeded $10^2 ~{\rm yen}/{\rm m}^2$.
 The number of data points is ``13,431''.}
 \label{78vs79}
\end{figure}

It is well known that the Pareto index for firm sizes hardly changes 
(Fig.~\ref{DistributionData}).
In such a database,
the detailed balance (\ref{Detailed balance}) is observed~\cite{FSAKA}
(Fig.~\ref{logProfit0405DB}).
On the other hand, the Pareto index for the assessed value of land varies annually.
This suggests that the system is not stable and the detailed balance does not hold.
Actually most of the period, the detailed balance is not observed in the scatter plot of
all pieces of land assessed in the database 
(Figs.~\ref{95vs96} and \ref{06vs07} for instance).\footnote{
Of course, the detailed balance is observed approximately
in the case that the system is almost stable and the Pareto index hardly varies
(Fig.~\ref{78vs79} for instance).}

\begin{figure}[htb]
 \centerline{\epsfxsize=0.75\textwidth\epsfbox{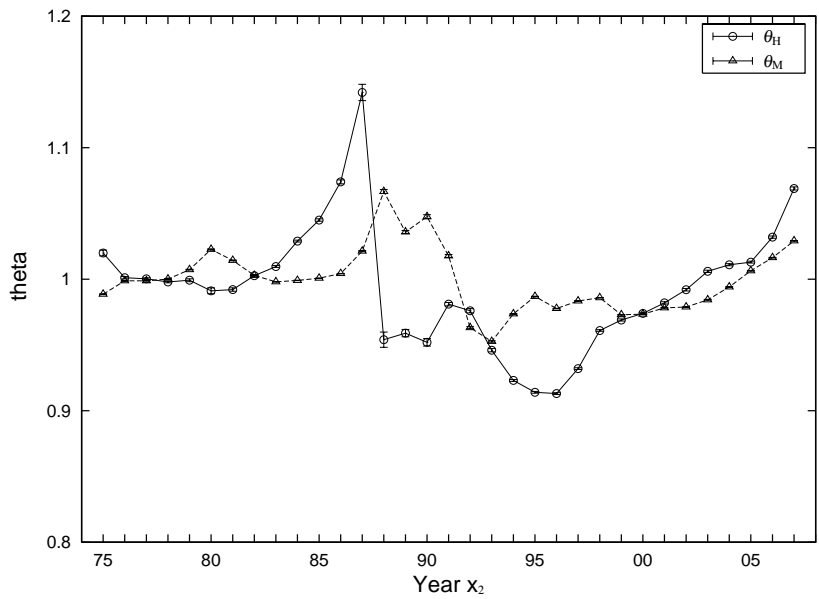}}
 \caption{Annual change of $\theta_{\rm H}$ and $\theta_{\rm M}$ in
 the year $(x_1, x_2) = (1974, 1975)$ -- $(2006, 2007)$.
 }
 \label{DqB}
\end{figure}

In the previous section,
we have proposed the detailed quasi-balance (\ref{Detailed quasi-balance})
in an ideal quasistatic system.
In each scatter plot of all pieces of land assessed in the database, 
the parameter $\theta$ is measured and the result
is shown in Fig.~\ref{DqB}.
Here $\theta$ is estimated in the following two regions.
One is the large scale region between $2 \times 10^5$ and $10^7$ ${\rm yen}/{\rm m}^2$
where Pareto index $\mu$ is estimated. The other is the middle scale one between
$5 \times 10^3$ and $3.17 \times 10^5$ ${\rm yen}/{\rm m}^2$
where the standard deviation $\sigma$ of the log-normal distribution is estimated.
In Fig.~\ref{DqB},
we represent two parameters by $\theta_{\rm H}$ and $\theta_{\rm M}$, respectively.

\begin{figure}[htb]
 \centerline{\epsfxsize=0.75\textwidth\epsfbox{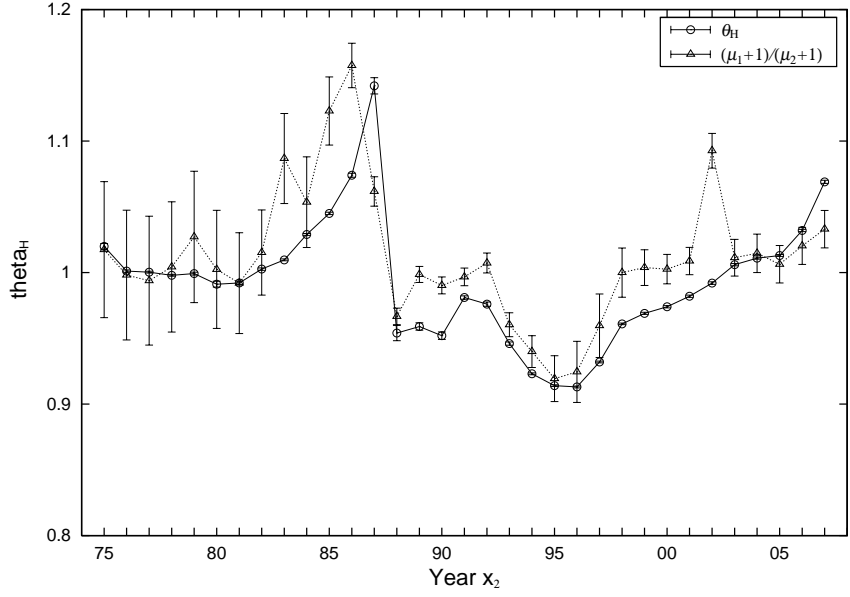}}
 \caption{Annual change of $\theta_{\rm H}$ and $(\mu_1+1)/(\mu_2+1)$
 in the year $(x_1, x_2) = (1974, 1975)$ -- $(2006, 2007)$.
}
 \label{Ratio}
\end{figure}

\begin{figure}[htb]
 \centerline{\epsfxsize=0.75\textwidth\epsfbox{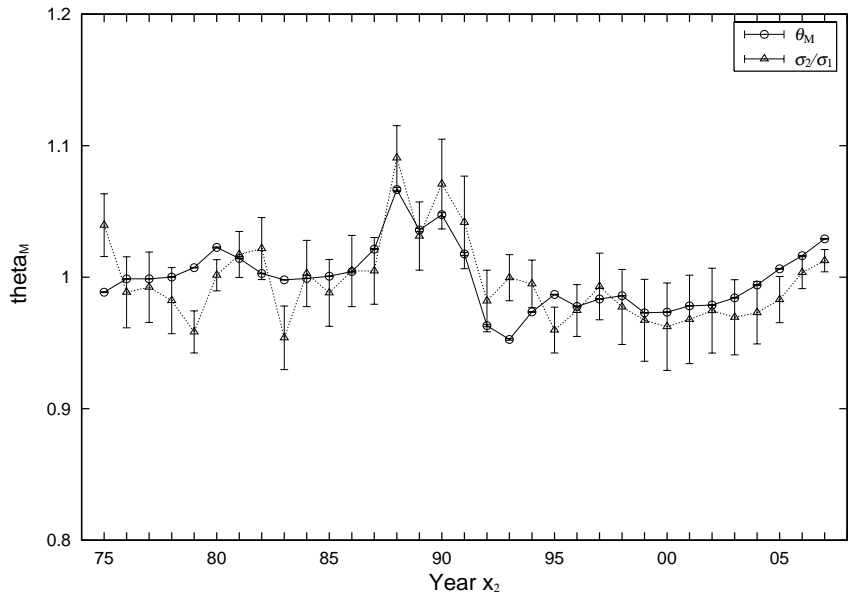}}
 \caption{Annual change of $\theta_{\rm M}$ and $\sigma_2/\sigma_1$
 in the year $(x_1, x_2) = (1974, 1975)$ -- $(2006, 2007)$.
}
 \label{Ratio2}
\end{figure}

By using these parameters,
we confirm not only the relation 
between the ratio of $\mu+1$ and $\theta_{\rm H}$
(\ref{ratio1}) in the large scale region
but also the relation 
between the ratio of $\sigma$ and $\theta_{\rm M}$ (\ref{ratio2})
in the middle scale one (Figs.~\ref{Ratio} and \ref{Ratio2}).
This warrants approximations assumed 
in the previous section.
The derivation and the data analysis are consistent.

\section{Conclusion}
\label{sec-Conclusion}
\indent

In this study,
by using no model, 
we have derived a quasistatically varying log-normal distribution 
from a Non-Gibrat's law 
under the detailed quasi-balance.
In the derivation, we have 
employed 
two approximations.
One is that the probability density function of the growth rate is described as tent-shaped
exponential functions.
The other is that the value of the origin of the growth rate distribution is constant.
Even under the detailed quasi-balance,
the first approximation uniquely fixes a Non-Gibrat's law to be the same expression
under the detailed balance.
Together with the second approximation,
the resultant distribution is described as power-law with varying Pareto index 
in the large scale region.
In the middle scale region, the distribution is reduced to
the quasistatic log-normal distribution with the varying standard deviation.
Notice that not only the change of Pareto index $\mu$ 
but also the change of the standard deviation $\sigma$ 
depends on a parameter $\theta$ of the detailed quasi-balance.

Employing empirical data on the assessed value of land in 1974--2007 Japan,
we have confirmed these analytic results.
In the scatter plot of all pieces of land assessed in the database,
the parameter $\theta$ of the detailed quasi-balance
is measured in the following two regions.
One is the large scale region where Pareto index $\mu$ is estimated.
The other is the middle scale region where the standard deviation $\sigma$ is estimated.
We have observed that two parameters $\theta_{\rm H}$ measured in the large scale region
and $\theta_{\rm M}$ measured in the middle scale one 
are in good agreement with the ratio of $\mu+1$ and $\sigma$, respectively.

Intriguingly, 
it is observed that
the change of Pareto index influences the change of the standard deviation 
in the opposite direction (Fig.~\ref{VaryingParetoIndex2}).\footnote{
This relation is suggested in Ref.~\cite{Souma} 
by employing personal income data in Japan.}
In this database,
the change of the distribution
in the large scale region propagates in the middle scale one.
This phenomenon is intelligible by two slopes $\theta_{\rm H}$ 
and $\theta_{\rm M}$ of the regression lines in the scatter plot (Fig.~\ref{DqB}).

In order to make these discussions more precisely, we need to investigate
wealth of data in unstable state for a long period.
By analyzing the database, we directly observe the Non-Gibrat's law
which defines the large and middle scale regions.
In addition, we should take a statistical test for the symmetry in the two
arguments of $P_{12}(x_1, x_2)$ to confirm the detailed quasi-balance directly. 
These accurate data analyses are imperative to realize the applications
of the study in this paper, 
such as
credit risk management and so forth.

\section*{Acknowledgments}
\indent

The author is 
grateful to 
the Yukawa Institute for Theoretical 
Physics at Kyoto University,
where this work was initiated during the YITP-W-05-07 on
``Econophysics II -- Physics-based approach to Economic and
Social phenomena --'',
and especially to 
Professor~H. Aoyama for the critical question
about my previous work.
Thanks are also due to  Dr.~Y. Fujiwara, Dr.~W. Souma and Dr.~M. Tomoyose
for a lot of useful discussions and comments.



\newpage

\end{document}